# Dehydrogenation through pressure-induced polymerization of phosphine

Ye Yuan[#,1], Yinwei Li[#,2], Guoyong Fang,[#,3] Guangtao Liu[1], Cuiying Pei[1], Xin Li[1,4], Haiyan Zheng[1], Yuexiao Pan[3], Ke Yang[5], and Lin Wang[*,1]

[1]Center for High Pressure Science and Technology Advanced Research (HPSTAR), Shanghai 201203, China

[2]School of Physics and Electronic Engineering, Jiangsu Normal University, Xuzhou 221116, China

[3]Key Laboratory of Carbon Materials of Zhejiang Province, College of Chemistry and Materials Engineering, Wenzhou University, Wenzhou 325035, China

[4]Department of Physics, Fudan University, Shanghai 200433, China

[5]Shanghai Institute of Applied Physics, Chinese Academy of Sciences, Shanghai 201203, People's Republic of China

[#]Y. Y., Y. L. and G. F. contribute equally to this work

[*]To whom correspondence may be addressed. Email: wanglin@hpstar.ac.cn

$PH_3$ is studied to understand its superconducting transition and stoichiometry under high pressure using Raman, IR, and x-ray diffraction (XRD) measurements, as well as theoretical calculations. $PH_3$ is stable below 11.7 GPa and it then starts to dehydrogenate through two dimerization processes at room temperature and pressures up to 25 GPa. Two resulting phosphorus hydrides, $P_2H_4$ and $P_4H_6$, were verified experimentally and can be recovered to ambient pressure. On further compression above 35 GPa, the $P_4H_6$ directly decomposes into elemental phosphorus. The superconductivity transition temperatures of $P_4H_6$ at 200 GPa is predicted to be 67 K, which agrees with reported result suggesting that it might be responsible for superconductivity at higher pressures. Our results clearly show that $P_2H_4$ and $P_4H_6$ are the only stable P-H compounds between $PH_3$ and elemental phosphorus, shedding light on the superconducting mechanism.

Since superconducting mercury was first reported [1,2], scientists have continued searching for new high-$T_c$ materials. In 2004, Ashcroft studied hydrogen-dominant hydrides [3], in which condensed $H_2$ may contribute to a high $T_c$. Motivated by this work, extensive theoretical investigations on this system have been reported, such as $SiH_4$ [4], $GeH_4$ [5], $GaH_3$ [6], $SiH_4(H_2)_2$ [7], $CaH_6$ [8], $YH_6$ [9] *etc*. However, a few remarkable high-$T_c$ materials were observed in subsequent experimental studies. Recently, Drozdov et al. reported the superconductive transition of $H_2S$ at 203 K at 155 GPa [10], which broke the highest $T_c$ record [11]. Many theoretical [12,13] and experimental [14] studies have explored its stoichiometry and structure, which play an important role in understanding the underlying mechanism of superconductivity.

Very recently, $PH_3$, a typical hydrogen-rich hydride, has attracted a great deal of research interest because of its superconducting transition by Drozdov and his co-workers [15–20]. Their experimental work revealed that $PH_3$ might be a high-temperature superconducting candidate. From the resistance measurements, a signature of superconducting transition was observed at the critical temperature ($T_c$) of 30 K. This increased to 103 K with pressures up to 207 GPa. However, structural information was not provided, and the origin of the superconducting transition remains puzzling. Subsequent theoretical studies [16–19] showed that the P-H compound should also be a complex system, and all the predicted structures were metastable with respect to the elemental phase.

Flores-Livas et al. [16] studied the phase diagram of phosphorus hydrides with different stoichiometries and found that phosphorus hydrides tended to decompose into phosphorus and hydrogen at high pressure. Liu et al. [17] reported a $PH_3$ phase with monoclinic structure (C2/m) and a $T_c$ of 83 K at 200 GPa, which is closer to the observed superconducting transition temperature. Shamp et al. [18] predicted that $PH_3$ is thermodynamically unstable during decomposition into the elemental phases, as well as $PH_2$ and $H_2$. Two $PH_2$ phases with C2/m and I4/mmm symmetry were computed as metastable at 200 GPa. The corresponding superconducting critical temperatures were 76 and 70 K, respectively. Bi et al. [19] found that a dynamically stable $PH_2$ phase was the best according to the observed superconducting transition at 80 GPa. The $PH_3$ phase to $PH_2$ phase reaction was exothermic at that pressure, which proves the spontaneity of the reaction.

To date, the $PH_3$ phase under compression remains unknown. No relevant experimental studies have been reported. The high-pressure stoichiometry and structural behavior of $PH_3$ are critical to understanding the superconducting transition in the P-H system, which needs to be experimentally determined. For this purpose, we studied the structural behavior of $PH_3$ under high pressure. We identified the pressure-induced step-by-step polymerization of $PH_3$ and a route to elemental phosphorus that unveiled the unknown transition process and provides experimental evidence for understanding the underlying mechanism of superconductivity of P-H compounds.

Solidified $PH_3$ was prepared *via* a cryogenic method and sealed into a diamond anvil cell (DAC) at ~ 2 GPa for *in-situ* high-pressure measurements. Raman spectra of the sample used a micro-Raman system (Renishaw, UK) with 532 nm laser excitation.

The high-pressure *in-situ* IR spectra were collected on a Bruker VERTEX 70v FTIR spectrometer and a custom-built IR microscope. High-pressure XRD measurements were carried out at the BL15U1 beam line of the Shanghai Synchrotron Radiation Facility ($\lambda$ = 0.6199 Å) [21]. The *ab* initio structure predictions for $P_4H_6$ were performed using the particle swarm optimization technique implemented in the CALYPSO code [22,23]. CALYPSO has been used to investigate many materials at high pressures [24–26]. The *ab* initio structure relaxations were performed using density functional theory (DFT) with the Perdew-Burke-Ernzerhof (PBE) generalized gradient approximation (GGA) implemented in the Vienna *ab initio* simulation package (VASP) [27]. Details of the simulations are provided in the Supplemental Material.

After the $PH_3$ gas was loaded into the sample chamber of a DAC and returned to room temperature, a colorless and transparent sample (Fig. S1) was observed. The characteristic Raman peaks (Fig. S1) at 978 ($\upsilon_2$, symmetric bending mode), 1104 ($\upsilon_4$, asymmetric bending mode), 2317 ($\upsilon_1$, stretching mode), and 2331 ($\upsilon_3$, stretching mode; shoulder) cm$^{-1}$ agreed well with previous reports [28], indicating the existence of $PH_3$ in the chamber.

The x-rays can damage the sample (Fig. S2), and Raman and IR are mainly used for *in-situ* studies of $PH_3$ at high pressure. Figure 1 (a) and (c) show the Raman spectra of the sample during compression. Under high pressure, these characteristic modes were blue shifted and broadened (Fig. 1 (b) and (d)). This eventually vanished at 20.5 GPa. Several new peaks (marked by black asterisks and arrows in Fig. 1 (a)) were observed at around 11.7 GPa, which suggested a phase transition. For the P-H stretching modes, we also noticed a dramatic expansion of the characteristic bonds. Figure 1 (d) shows the peak positions of the $\upsilon_1$ and $\upsilon_3$ modes as a function of pressure. The peak shift of $\upsilon_1$ dramatically decreased and started to red shift at 11.7 GPa. We attributed these changes to a transition in the sample near 11.7 GPa.

These new peaks in the Raman spectra (Figure 1 (a)) were consistent with previous studies about $P_2H_4$ at ambient pressure. The two new peaks at low frequencies correspond to the $PH_2$ rocking mode and P-P stretching mode in the $P_2H_4$ molecule, which were observed at around 217 cm$^{-1}$ and 436 cm$^{-1}$, respectively [29,30]. The emergency P-P bond at 11.7 GPa proved the dimerization of $PH_3$ molecules. The other new peaks at 1007 and 1093 cm$^{-1}$ are from the $PH_2$ scissoring modes in $P_2H_4$ molecule, which also agreed with previous reports. These suggest that the pressure-induced transition is due to the dimerization of $PH_3$ at high pressure.

To verify the dimerization, we also studied the decompressed sample. A liquid sample is obtained after quenching to ambient conditions as seen from the microphotograph of the decompressed sample. It is well known that $P_2H_4$ is a liquid at ambient pressure [29,31], which confirms that pressure drives the dimerization of $PH_3$ to form $P_2H_4$ *via* this reaction:

$$2PH_3 \rightarrow P_2H_4 + H_2 \qquad (1)$$

We further employed Raman to measure the recovered liquid sample. However, after laser irradiation, the liquid sample decomposed and generated Hittorf's 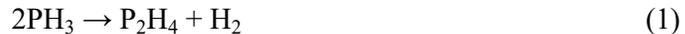

phosphorus [32,33] (Figure 2) according to the photodecomposition properties of $P_2H_4$ [31]. This offers more evidence of our findings.

We also employed infrared spectroscopy to trace the *in-situ* information of the new product at high pressure. Figure S3 (a) shows the IR peak near 1095cm$^{-1}$ broadened and shifted slightly to a lower frequency with increasing pressure, but an obvious new shoulder was observed at around 1058cm$^{-1}$ after decompressing the sample to 11.8 GPa (Fig. S3 (c) and (d)). This new shoulder matched the $P_2H_4$ scissors mode well, which was observed at around 1052 cm$^{-1}$ in a solid state at ambient pressure [34]. This characteristic mode confirms the existence of $P_2H_4$. In addition to the P-H stretching modes in the IR spectra (Fig. S3 (b)), a new shoulder at around 2329 cm$^{-1}$ was observed, and it became stronger and stronger with increasing pressure. After it had quenched to 11.8 GPa, the new shoulder peak was more obvious compared to the IR spectrum measured at 12 GPa during compression. This proves dimerization.

As the pressure increased, the $P_2H_4$ shows piezochromism. It becomes yellow, then red and darkened and eventually becomes opaque at pressures higher than 25 GPa consistent with the observations of Drozdov et al. at low temperatures (180 K). As the sample became totally opaque, the vibrational signal vanished, and hindered the *in situ* high-pressure vibrational spectra measurement. Therefore, we had to quench the sample to ambient conditions from different pressures (25 and 35 GPa) and employed Raman spectroscopy to investigate the different quenched residues. Interestingly, once the sample became completely opaque above 25 GPa, it maintained its opaque solid state even when decompressed to room pressure. This irreversible process suggests that a new transition occurred at higher pressures.

Figure 3 (a) shows that the Raman spectrum of the residue quenched from 25 GPa after the opaque transition. A weak peak near 873 cm$^{-1}$ belonging to $PH_2$ twisting and a strong peak at 2248.5 cm$^{-1}$ belonging to P-H stretching exist in the spectrum. This new P-H stretching peak is located at a much lower wavenumber than that from $PH_3$, $P_2H_4$ (~2292 cm$^{-1}$), and $P_3H_5$ (~2267 cm$^{-1}$) [31] suggesting that the residue contained a new kind of phosphorus hydride. Figure 3 (b) shows the P-H stretching mode of $P_nH_{n+2}$ shifts to lower frequency as n becomes larger. Following this trend, we deduced that the new phosphorus hydride was $P_4H_6$, which suggests that the $P_2H_4$ molecules continued to dimerize and form $P_4H_6$ at higher pressure.

To confirm the second dimerization, we calculated the Raman modes of $P_4H_6$ using Gaussian 09 program at the B3LYP/6-311(d,p) level [35]. Table S1 in the supplemental material shows the calculated Raman modes of two typical $P_4H_6$ conformers in which four phosphorus atoms are linear and U-type (Fig. S4 (a) and (b)). The calculated Raman spectra show that they both have four characteristic bands corresponding to the stretching vibration (350~450 cm$^{-1}$) of the P-P bond, twisting vibration (700~900 cm$^{-1}$) of $PH_2$ group, scissoring vibration (ca. 1070 cm$^{-1}$) of $PH_2$ group, and stretching vibration of P-H bond, respectively. Moreover, the P-H stretching mode can further shift to a lower frequency (2278 cm$^{-1}$). From Table S1, we can see that the P-P stretching bonds and the twisting vibration of the $PH_2$ group from linear $P_4H_6$ are closer to our observed peak suggesting that the linear type $P_4H_6$ is the more possible conformer in the residue.

Besides the peaks from $P_4H_6$, several obvious characteristic modes (123.8, 184.8, 218.9, 285, 357.2, 386.5, 407.7, 443.2 and 505.8 cm$^{-1}$) were observed below 550 cm$^{-1}$. These peaks are similar to fibrous red phosphorus characteristic modes [32,33], which indicated that parts of $P_4H_6$ was thoroughly dissociated when exposed to laser or decompression. At ambient pressure, phosphorus hydrides often undergo disproportionation into phosphorus-rich phosphanes upon exposure to light and heat [31]. For example, $P_2H_4$ can decompose into $P_3H_5$, and $PH_3$, $P_3H_5$ can decompose into $P_4H_6$ and $P_2H_4$. However, we did not observe the Raman peaks from $P_3H_5$ or other phosphanes from the residue, which proves that $P_2H_4$ dimerized directly into $P_4H_6$ at high pressure corresponding to the equation as:

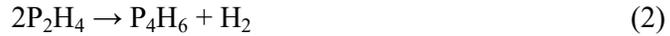

$$2P_2H_4 \rightarrow P_4H_6 + H_2 \qquad (2)$$

Figure 3 (c) shows that the Raman spectra of the residue quenched from 35 GPa. After decompression to 1 atm, typical black phosphorus modes were observed [36,37]. Therefore, $P_4H_6$ eventually decomposed into elemental phosphorus at 35 GPa. Hence, the corresponding reaction is as follows:

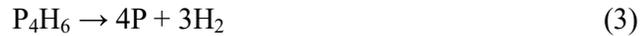

$$P_4H_6 \rightarrow 4P + 3H_2 \qquad (3)$$

From the *in-situ* high pressure XRD (Fig. 3 (c)), typical diffraction rings of cubic phosphorus further confirmed the thorough decomposition of $P_4H_6$ at high pressure.

The superconductivity of elemental phosphorus has already been studied both experimentally and theoretically [38–40]. The maximum $T_c$ is about 9.5 K at 32 GPa before decreasing with pressure. Near 100 GPa, the $T_c$ is about 4.3 K. At 160 GPa, and no superconducting transition was detected in the temperature range from 4 to 40 K. The much lower $T_c$ of phosphorus compared to 100 K indicates that $PH_3$ or other phosphorus hydrides should be responsible for the superconductivity observed at 200 GPa in Drozdov's work. Our experiments show that $PH_3$ is not stable and goes two-steps polymerization. It eventually decomposes into phosphorus at high pressure suggesting two new phosphorus hydrides $P_2H_4$ and $P_4H_6$ are responsible for superconductivity.

The phase transition pressures in our studies are much lower than Drozdov's. We speculate that this discrepancy is due to the different experimental protocols used in these two works. In our experiments, the sample was compressed at room temperature, while Drozdov's increased the pressure at T < 200 K. [15] The temperature could highly reduce the phase transition pressure, as found in many other high pressure and high temperature syntheses, e.g., high temperature can effectively prompt graphite to diamond under high pressure. Furthermore, we observed similar phenomenon, indicating the samples went through the same phase transformation process. For instance, we found the Raman signal decreased and vanished when the sample darkened at around 25 GPa. It became opaque and began to conduct at about 30 GPa. There were no obvious color changes upon further compression. These piezochromism phenomena are similar to the pressure effect on $P_2H_4$ and $P_4H_6$, which suggested that $PH_3$ also underwent dimerizations in their work at low temperatures. We also realize that the pressure dependence of $T_c$ becomes slower near 150 GPa, which suggests that a transition occurred. As predicted by Shamp et al, the I4/mmm $P_2H_4$ has a $T_c$ of 50 K at 150 GPa, which agrees with the measurements very well, suggesting the superconducting phase is $P_2H_4$ at low pressure. $P_4H_6$ might be a superconducting component at higher pressures.

We further performed structural searches on $P_4H_6$ at 100, 150, and 200 GPa with maximum simulation cells up to 4 formula units (f.u.); two stable structures with space group Cmcm (< 182 GPa) and $C2/m$ (> 182 GPa) were found. Phonon dispersions calculations of the two structures do not give any imaginary frequencies and therefore this verifies their dynamic stabilities (Fig. 4). The superconducting $T_c$ was estimated using the Allen and Dynes modified McMillan equation [41] with a typical choice of $\mu^*$=0.13. The electron-phonon coupling constant λ of the Cmcm structure is only 0.59 (Table) at 100 GPa, and a superconducting $T_c$ of 13 K was obtained. A relatively large λ value of 1.39 was found for the C2/m structure at 200 GPa, and the superconducting $T_c$ was estimated to be 67 K. As summarized in Table 1, the estimated $T_c$ agree with the values measured by Drozdov et al. suggesting that $P_4H_6$ could be responsible for the superconductivity.

Similar to $H_2S$, $PH_3$ is unstable at high pressure. Instead of becoming more hydrogen enriching, it dehydrogenates through a series of polymerization/decomposition processes upon compression. This could be one of the critical facts that limit the maximum $T_c$ near 100 K at the same pressure where H-S system has a $T_c$ up to 180 K. Inspired by these phenomena from $H_2S$ and $PH_3$, avoiding the pressure-induced dehydrogenation or becoming more hydrogen enriched is a key issue to consider when looking for a superconducting hydride with a high $T_c$.

In summary, we determined the stability of $PH_3$ under high pressure. Two steps of polymerization were obtained. $P_2H_4$ and $P_4H_6$ were the reaction products of the first and second step dimerization, respectively. Above 35 GPa, the generated $P_4H_6$ completely decomposed into elemental phosphorus. Vibrational measurements and theoretical simulation confirmed the formation of $P_2H_4$ and $P_4H_6$, which enriches the phase diagram of the P-H system under high pressure. Our work proves the generation of $P_2H_4$ phase under high pressure and suggests that it might be responsible for the superconductivity observed at pressures below 150 GPa. The $P_4H_6$ could be the superconducting phase at higher pressures.

## Acknowledgement

The authors acknowledge the support of NSAF (Grant No: U1530402) and Science Challenging Program (Grant No. JCKY2016212A501).

# Figures

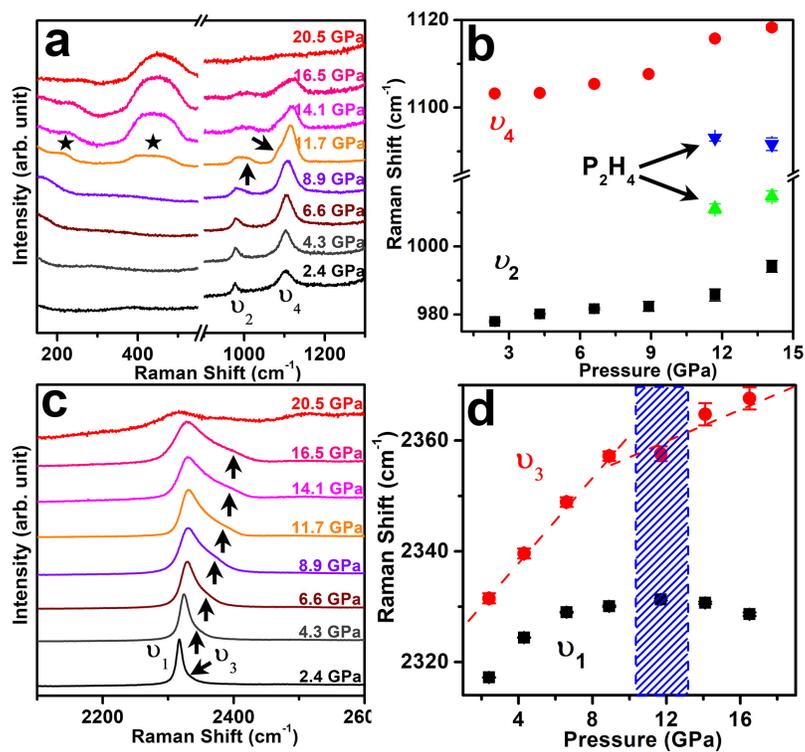

FIG. 1. (a) and (c) Raman spectra of $PH_3$ collected at various pressures at room temperature. The peak positions of $\upsilon_2$, $\upsilon_4$ (b) and $\upsilon_1$ $\upsilon_3$ (d) as functions of pressure, the dash line is a guide for eyes.

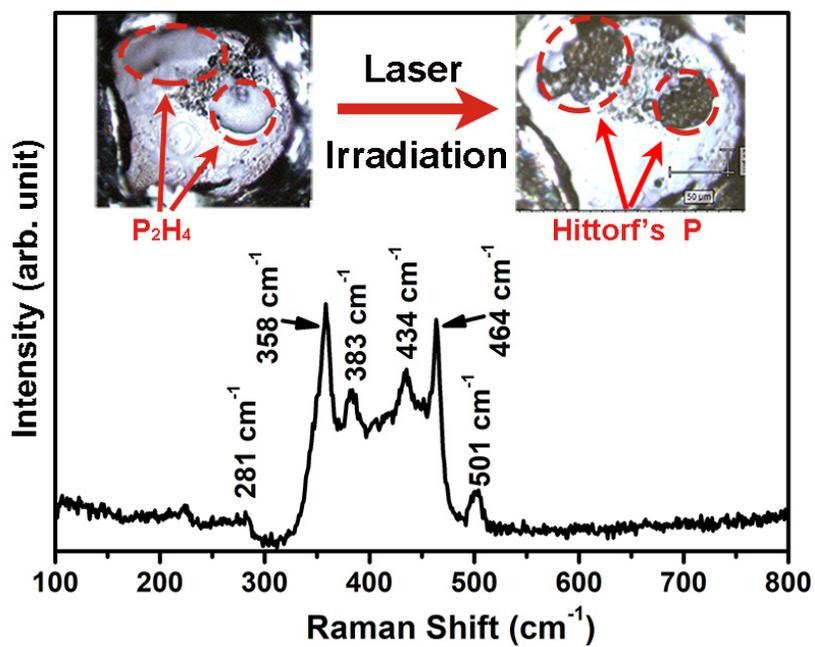

FIG. 2. The Raman spectrum of the Hittorf's phosphorus transformed from the liquid sample after laser irradiation. The inset optical images show the photo-induced transition of the liquid residue before and after laser irradiation.

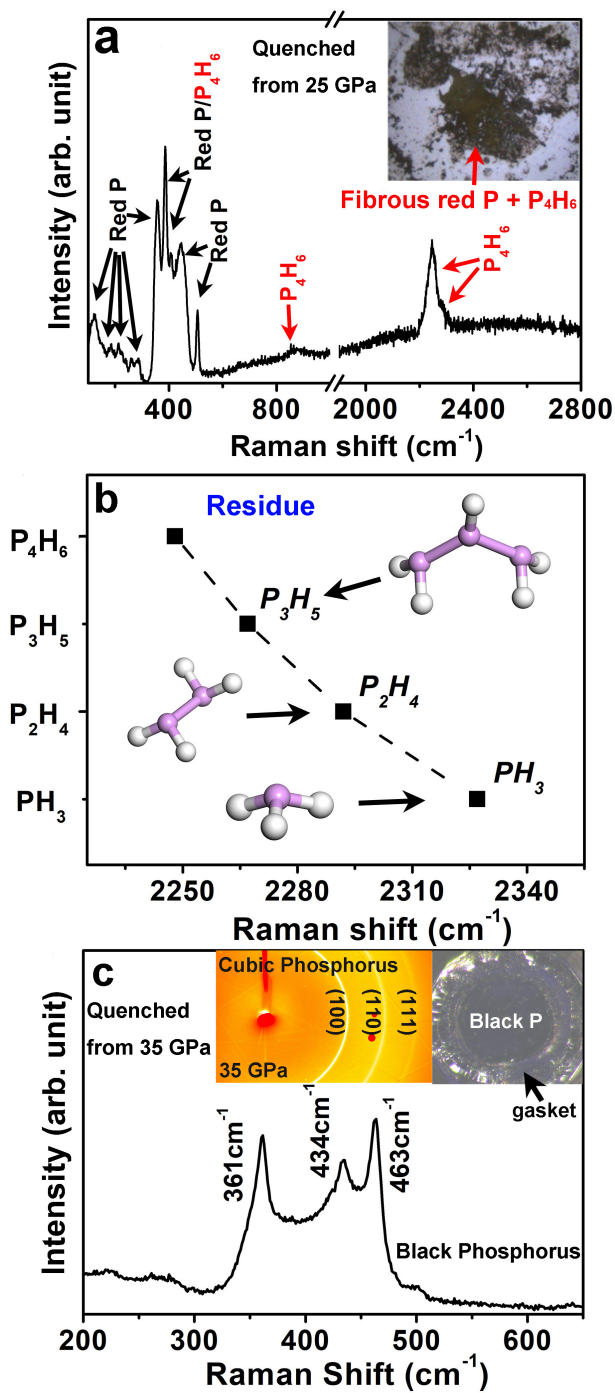

FIG. 3. (a) The Raman spectrum of the sample decompressed from 25 GPa, showing the residue is the mixture of fibrous red phosphorus and $P_4H_6$. The inset picture shows the optical micrograph of the decompressed sample. (b) The tendency of frenquency of the P-H stretching in $P_nH_{n+2}$ (n=1, 2, 3, and 4). (c) The Raman spectrum of the sample quenched from 35 GPa. The inset show the XRD pattern of the sample at 35 GPa and optical micrograph of the sample released to 1 atm.

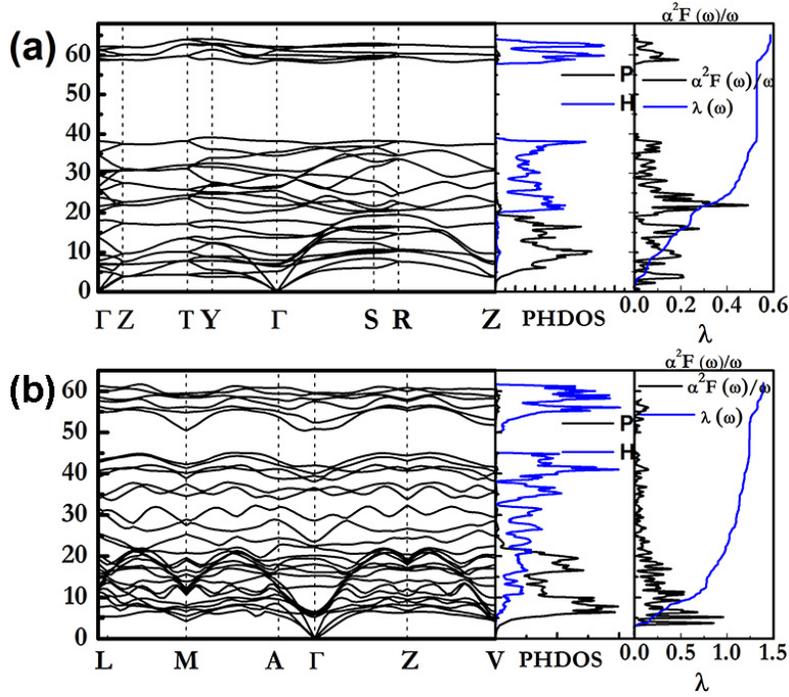

FIG. 4. Phonon dispersions, phonon density of states projected onto atoms (PHDOS), the spectral functions $\alpha^2F(\omega)/\omega$ and electron-phonon coupling integration of $\lambda(\omega)$ for (a) Cmcm structure at 100 GPa and (b) C2/m structure at 200 GPa, respectively.

Table The calculated electron-phonon coupling constants ($\lambda$), the logarithmic average phonon frequency ($\omega_{\log}$), and $T_c$ with $\mu^*=0.13$.

| Phases | Pressure (GPa) | $\lambda$ | $\omega_{\log}$ | $T_c$ ($\mu^*=0.13$) |
|---|---|---|---|---|
| Cmcm | 100 | 0.59 | 889 | 13 K |
| C2m | 200 | 1.39 | 700 | 67 K |